\documentclass{cs20proc}

\usepackage{hyperref}

\editors{S. J. Wolk, A. K. Dupree, H. M. G\"unther}
\publisher{Zenodo}
\conference{The 20.5th Cambridge Workshop on Cool Stars, Stellar Systems, and the Sun}
\conferencedate{2021}
\title{How to organize an online conference -- Lessons learned from Cool Stars 20.5 (virtually cool)}

\author{Hans Moritz G{\"u}nther$^1$, James R. A. Davenport$^2$, Scott Wolk$^3$, Shaun Gallagher$^3$}
\affiliation{$^{1}$ MIT Kavli Institute for Astrophysics and Space Research, 77 Massachusetts Avenue, Cambridge, MA 02139, USA\\
$^2$University of Washington, Department of Astronomy, Seattle, WA 98195, USA\\
$^3$Center for Astrophysics | Harvard \& Smithsonian, 60 Garden Street, Cambridge, MA 02138, USA}

\shorttitle{How to organize an online conference}
\shortauthors{Cool Stars 20.5 LOC}

\abs{The virtual meeting was a success. Several people told us that this was “the best virtual meeting they had seen so far”, which, a year into the pandemic and without a commercial provider in the back, is a great success.
The biggest point of criticism was the timing: We had programming from UTC 17:00-22:00 (evening and night in central Europe, afternoon on the US East Coast, during the day in South America and on the US West coast, but in the middle of the night for Asia and Australia). There is no good solution, but at least some variation in session time might go a long way to make it easier for all to attend at least some sessions. Feedback also indicates that the schedule was too compressed.
Poster sessions and social contacts with the tool Gathertown worked out really well for all that used it.
Our way of combining several services (Zoom for plenary and break-out rooms, Zenodo for uploading and viewing posters and proceedings, Google forms for registration and abstract submission, gathertown) allowed for a very low-cost meeting with little overhead (total cost: 600 \$ for gathertown, zoom was provided through an institutional subscription, just 4 people on the LOC).
}

\begin{document}

\maketitle

\section{Setting}
Cool Stars 20 took place in person in Boston, MA, USA in the summer of 2018. Because of the pandemic, Cool Stars 21, scheduled for Toulouse, France in June 2020, had to be pushed back. The SOC for CS21 met in September 2020 and realized that meeting in summer 2021 was not viable. It was decided to organize a virtual interim meeting. The organization of CS 20.5 was chosen to leverage the organization of CS20 led by Scott Wolk (SOC chair) and Moritz Günther (LOC chair). Early March 2021 was chosen to meet as soon as practical (as we were already “late”) and to avoid summer and spring conflicts. The concept was to have real Cool Stars meeting, but keep the timing, organization, and costs lean. Our expectations were for a relatively small, interim, meeting so we expected about 9 hours of meeting time to be sufficient.

\section{Format and attendance}
The format was a daily synchronous sessions via a Zoom meeting. These sessions were 3 - 4 hours long (with breaks), and held from ~17:00 to 21:00 UTC. The meetings were recorded, and uploaded to the Cool Stars 20 YouTube channel\footnote{\url{https://www.youtube.com/channel/UCb5YV1cMVuvl2SIu33G02rg}} after the conference. The sessions included a combined five invited talks of 25 + 5 min (vs. 12 at CS 20), which were held live on zoom. We also had 29 contributed talks of 12 + 3 min (vs 39 at CS 20), also live on zoom, and about 50 ``haikus'' (1 min presentations) which were pre-recorded and merged into a video which was played during live sessions.

The largest in-person Cool Stars conference had about 500 attendees. CS 20.5 had 1200 online registrations, about 400 abstracts, and between 350 and 550 people on zoom at any given time. Some people registered more than once (see registration discussion below) and some attended only specific zoom sessions.

\begin{figure*}
	\centering
	\includegraphics[width=0.49\linewidth]{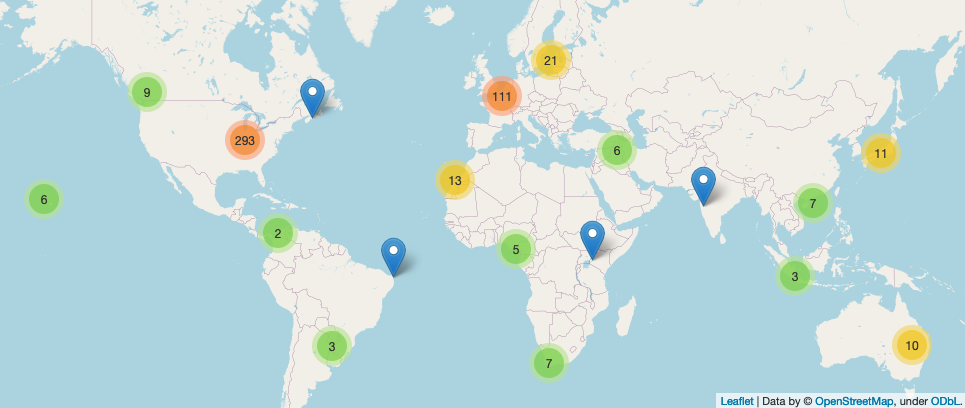}
        \includegraphics[width=0.49\linewidth]{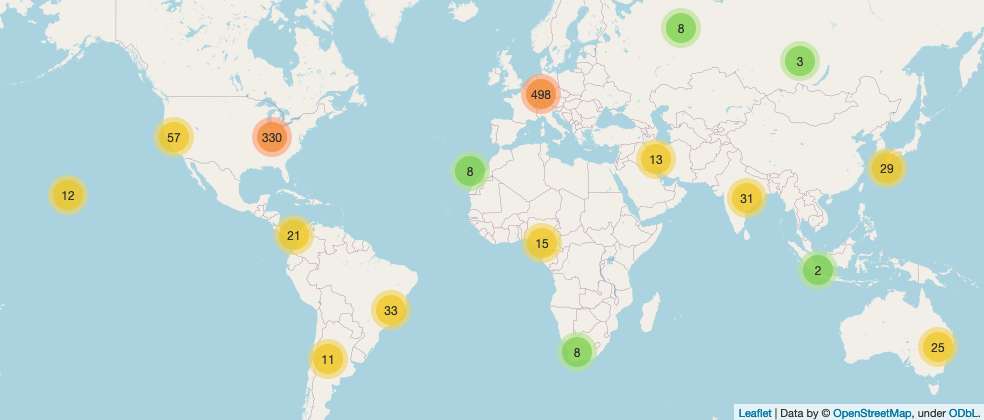}
	\caption{Location of home institutions for conference participants. \emph{left:} Cool Stars 20 in person in Boston, MA, USA, \emph{right:} Cool Stars 20.5, virtually. Interactive maps are available at \url{http://coolstars20.cfa.harvard.edu/cs20half/participants.html}.}
	\label{fig:attendance}
\end{figure*}

\subsection{Geographic diversity}
Figure~\ref{fig:attendance} shows a map with participants home institutions marked. Matching is done through automated query of bing maps and may not be correct in all cases (e.g.\ two participants listed their affiliation as "astronom", which gets mapped to an address in Denmark). Participants who did not provide an affiliation are located based on their email address (e.g. an address ending on ".de" will place a marker somewhere in Germany). It is obvious that virtual programming opens up Cool Stars to a much wider community. The number of participants from North America is similar for CS20 and CS20.5 indicating that essentially all interested researchers did travel to Boston for the in-person conference. However, there are four times more participants from Europe, which shows that even for institutions in relatively rich countries, long-distance travel is a hurdle, because of limited travel funds or time limitations (e.g. teaching or child-care). Although the total numbers are lower, attendance from South America, Asia (in particular India) and Africa is also much higher in the virtual event.

\subsection{Diversity by career stage}
Although we did not perform a quantitative survey, we know from feedback received after the conference, that the virtual programming allowed for the participation of, in particular, undergrad students and PhD students, who would not have been able to finance a trip to an in-person conference.

\subsection{Climate impact}
The carbon footprint of international in-person meetings is dominated by emission from long-distance flying. \citet{2020NatAs...4..823B} compared an in-person and a virtual meeting for a large astronomy conference, and found that a virtual meeting reduced the carbon footprint by 3-4 orders of magnitude. While we did not tally up individual flights, a look at Figure~\ref{fig:attendance} shows that attendees for Cool Stars are at least much, if not more, distributed over the world as the conference analyzed by \citet{2020NatAs...4..823B}. Thus, the carbon savings of holding CS 20.5 virtually, is at least 3-4 orders of magnitude compared to an in-person conference.

\section{Timeline and deadlines}
A requirement on the SOC was to organize CS 20.5 quickly. As such, the timeline for finding speakers, abstract submission etc.\ was extremely compressed. There were 12 weeks from inception to the abstract due date, which we felt was the minimum for reasonable publicity.  This left 4 weeks for the SOC to select speakers from over 350 abstracts. That’s a lot to read. Small subcommittees were formed so that no one had to grade more than 80 abstracts. The initial selection worked, but there is always some balancing needed in terms of speaker gender, topic, career stage, geographic origin etc, that leads to moving speakers around the program.  The compressed timeline meant that the SOC chair had to make some of those as executive decisions.  This is where an extra two weeks of time would have allowed the SOC a full say.

Similarly, we accepted poster abstracts till one week before the conference, and then required upload of posters on the Friday before the conference started. The timeline did not allow for the SOC to properly review any late posters AND give the presenters time to actually write their posters. Instead, we switched to “Upload your poster by the Friday before the conference and we’ll accept it unless it’s obvious junk” because we did not want to move the deadline to an earlier date than already publicly announced. Both dates were too tight and the “upload poster even if we have not reviewed your abstract” process caused some confusion for later poster submissions. We updated the website with new posters multiple times a day starting the week before the conference till after the conference started.  We also generated a poster movie, combined and edited the haiku movies, all during the weekend before the meeting started.

Notably, this compressed timeline meant that there was no time to check that posters sent to the gathertown organizer were actually also submitted to Zenodo and approved by the SOC. As far as we noticed, only one person snuck in a poster like this, but we had no cross-referencing to verify that no one else did this. For this conference that turned out not to be a problem, as poster presenters usually have broad leeway to change title and abstract, but conferences that want to control posters more tightly need more time in this step.

\subsection{There is no good solution to session timing.}
The scheduling of sessions was the most criticised point in feedback we received. Given that attendees spanned 19 time zones, there likely is no good solution, but there may be some that are less bad than others.

The goal of the SOC in establishing CS 20.5 was to hold a quick, interim meeting and thus we had a very compressed daily schedule with just three days of programming for five hours each: Four hours of plenary session and one hour of topical interest rooms (more open topical discussion in smaller groups). We scheduled only a single 15 min break in that time and even that was shortened when the session run over time. More frequent breaks are needed, even in an online setting.
Furthermore, the timing (UTC 17:00-22:00, with social gathertown meetings extending beyond that) was convenient for the US, particularly on the East coast, but painfully late for many Europeans and in the middle of the night for participants from Asia and Australia. Based on feedback from the participants, a more stretched schedule (four or five days) with less programming per day would have been easier. Additionally, schedules for sessions could shift such that each time zone has at least some session that is at a convenient time.
We note that even within a timezone, preferences vary. For most, normal daytime hours work best, but e.g. parents of remote-schooled children might actually prefer evening or early night hours.
Similarly, opinions are split over several days in a week as in a traditional conference vs. one-day-per-week for a longer period of time. Attendees noted that it is easier to free up one day per week, but the “feeling” of a conference with frequent interactions with the same people and topics is lost; a one-day-per-week meeting has more similarity with a colloquium series where attendees choose to listen to a few select talks only, instead of listening to a broader range of topics that are part of the full conference program.

\begin{figure*}
	\centering
	\includegraphics[width=0.49\linewidth]{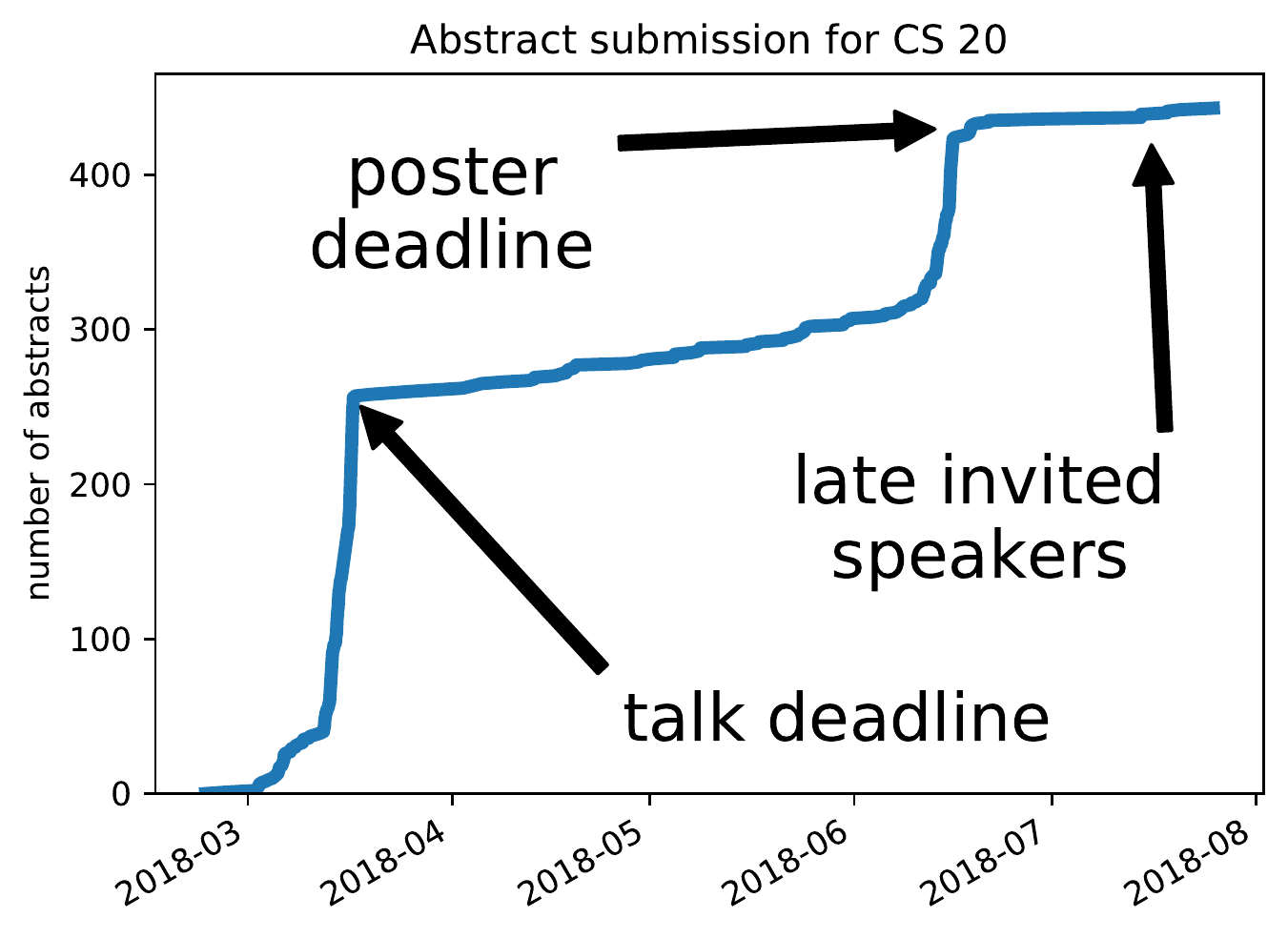}
        \includegraphics[width=0.49\linewidth]{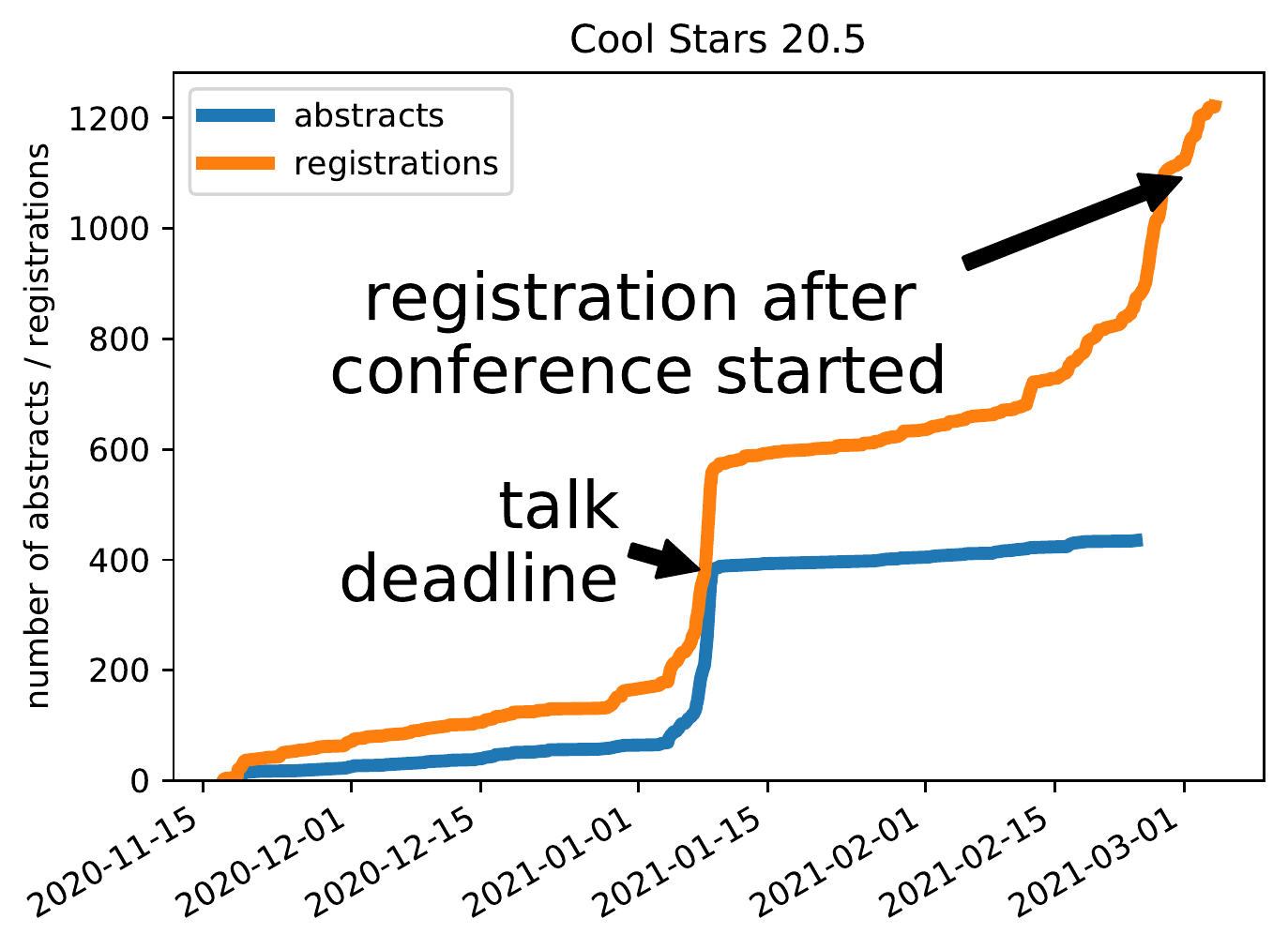}
	\caption{Timing of registration and abstract submission. \emph{left:} Cool Stars 20 in person in Boston, MA, USA (abstract submission), \emph{right:} Cool Stars 205., virtually (abstract submission and registration).}
	\label{fig:timeline}
\end{figure*}

\subsection{Workload}
For every conference, there are three phases of maximum workload for the LOC:
\begin{enumerate}
    \item Initial setup (chose venue for in-person conference, set-up website),
    \item abstract submission deadline (give abstracts to SOC in readable form),
    \item and
conference set-up and first day (in-person: Registration, virtual: start slack, finalize website, send out zoom links).
\end{enumerate}

For this virtual conference, we additionally had to organize, re-format, sort, and present the posters. In an in-person conference, attendees can be asked to pin up the posters themselves on poster-boards that a vendor provides. The problem is that in our timeline, the ``final handling of posters on our website and Zenodo'' coincided with the conference set-up. We recommend setting the poster upload deadline 10 days (and not just one weekend) before the beginning of the conference.

Figure~\ref{fig:timeline} shows the timeline of abstract submission for CS20 and CS 20.5 and the registration for CS20.5 (registration for CS20 was handled by a different system and we do not have time-stamps available now). This clearly highlights a tendency of astronomers to submit on (or slightly after) the deadline.

For conference organizers, this leads to the following rule-of-thumb: Don’t worry if you see very few abstracts submitted initially. At most 50\% of the abstracts are submitted more than a week before the deadline. To turn this argument around: There is no way to see what the interest in your conference is until a few days before the abstract deadline, so, if you need to adjust vendor contracts (e.g.\ zoom license, poster boards) based on number or abstracts / participants, make sure that your abstract deadline is before the date you need to know those numbers.

\section{Tools}   \label{sect:tools}
We now discuss the tools we used.
\subsection{Website: Static, but with scripts to generate some content}
\label{sec:website}
The website followed closely what we had written for CS 20. All relevant code is public on github (https://github.com/CoolStars20/websitehalf) and has been used for 6 major conferences now. The idea is to have static websites where most of the html is hand-written, but certain parts are generated from input data using Python scripts. Most importantly, the participant list and the abstracts and schedule are generated from a csv file that holds all the registration and submitted abstracts. Static websites are easy to host, because you just need to copy html files to a server, there is none of the work involved with setting up of SQL databases etc. On the other hand, abstracts and participant lists change frequently, so it’s very annoying and error-prone to edit the html by hand for all of those, hence the .csv processing tools.

This led to a mismatch. Participants expected to be added to the list of participants immediately after registration and contacted us when that was not the case. We explained this was a manual workflow where the LOC chair downloaded the csv file every now and then and then updated the page. Adding a sentence “This list is manually updated” solved the problem.

Similarly, we pulled the posters from Zenodo and placed them with thumbnail images on our website manually. Again, people expected them to show instantly.
One advantage of the manual approach is that it’s easy to manually check the website before upload and spot some obvious errors (e.g. a single person registering multiple times and thus appearing multiple times in the participant list) or code errors (e.g. sorting with non-ASCII characters in names). On the other hand, automatic updates using a script that is run every x hours (and more often on the day of the deadline), would push updates faster and save time in the critical few days before the conference. Of course, most of this work scales with the number of participants. The smaller the conference, the lower the benefit of automation.
What we did worked, but if we were to organize a conference with > 500 participants again, more automation would be added.
The website received little feedback beyond listed above, which means it did its job.

\subsection{Registration: Google forms}
\begin{figure*}
	\centering
	\includegraphics[width=0.49\linewidth]{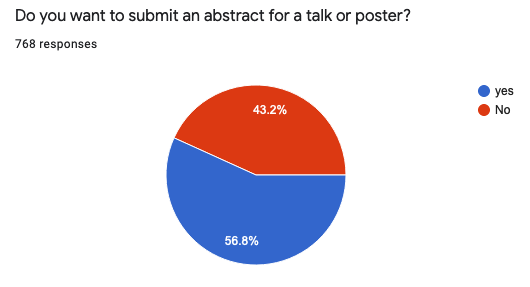}
        \includegraphics[width=0.49\linewidth]{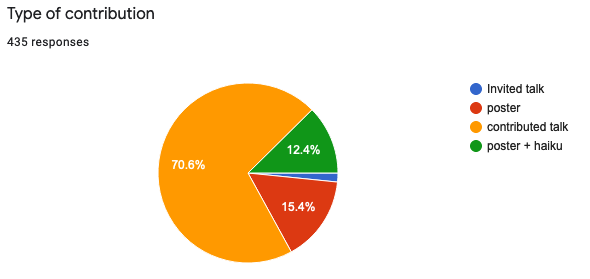}
	\caption{Two overview graphics from Google forms (screenshot from the Cool Stars 20.5 abstract submission form). \emph{left:} About one third of the participants submitted an abstract. \emph{right:} Talks are the most popular -- and most oversubscribed -- option.}
	\label{fig:googleforms}
\end{figure*}

Registration was done through a google form with no registration fee. The form was set to send a confirmation email, which a significant fraction of people did not get (it probably went to their spam), leading to questions per email ``Did my registration go through?'' or to people registering multiple times (with the same or different email addresses). Having a regularly updated list of participants on the website helped, as people checked for their name there (but see the point in section~\ref{sec:website} about website infrastructure).

We combined registration and optional abstract submission into a single form. In CS20, those were separate (registration went through a different site that processed credit cards) and it’s a lot of manual work to sync up two different systems because people use different email addresses, full first names or initials, non-ASCII characters in names are handled differently etc. Combining both in one form resolved these difficulties.

Google forms sends a confirmation email with a unique link that allows participants to edit their contribution until the form is switched inactive. We asked invited speakers to use the same form to submit their abstracts, which at the same time makes sure they are actually registered on the email list for participants (invited speakers tend to forget to register). Only 2/400 abstract submissions selected ``invited'' when, in fact, they were not. That is easily corrected by hand by editing the google sheet.

The LOC then passed the submitted abstracts to the SOC. There are several ways to display that information:
\begin{itemize}
\item In google forms, as pie-charts (number of registrations, abstracts per topic), see Fig~\ref{fig:googleforms} for examples.
\item In a google sheet that collects all answers (but long abstracts are hard to read in a spreadsheet).
\item Downloaded as .csv and processed with a script to format as html or latex/pdf.
\end{itemize}
Most of the SOC used the pdf version, generated from the .csv table, to read abstracts. We had 350 abstracts from which to choose $<30$ contributed talks and 60 1 minute ``haiku''.
A consistent numbering is important to keep track. The SOC selected voted on contributions using excel, a unique identifier was important to track each individual contribution. The final result was that contributions were assigned time slots which the LOC manually entered additional columns in the google sheet. That was clumsy, but satisfactory, because it only needs to be done once.

For CS20, we ran a script that checked the Google form every few minutes and sent out a notification to our own email account. That worked well and should have been done here, too. In particular, we still received $> 100$ registrations within 24 hours of the conference start(some even after that). For the late registrations, we had to manually send out emails with the zoom links, slack invite link etc. If we had set up an automated bot, we could have placed all this information in the automated response and participants would have received it within minutes of registering without us doing any work in a phase of the conference when the organizers are busy already.

None of our registrations were obviously fake or from a spam account. Apparently Google forms is very good at blocking that or our conference is not prominent enough to attract trolls.

We used a commercial email marketing provider (a free account on mailchimp) to announce the conference to a list of previously collected email addresses (from previous Cool Stars meetings) and also to distribute information (e.g. the zoom links) to registered attendees. Mailchimp allows us to create groups, one group was “CS20.5 registered”, only these people received the final meeting links. This worked well for most cases and is relatively easy to set up. The mailchimp software takes care of unsubscribe requests, etc. However, for some fraction of attendees, the mail went to spam and we had to manually forward the information. We do not think there is a better solution - we emailed the speakers from our personal accounts and some of that information was also lost to spam filters. The lesson to take from that is to have multiple methods of communication during the conference (email, slack, website) and to monitor the conference organizer email closely starting 24 h before and during the conference, which we did.

\subsection{Poster and Haiku upload: Zenodo}
We used Zenodo\footnote{\url{https://zenodo.org}} to collect posters and haikus. We chose Zenodo, because it is an EU-funded long-term archive that assigns DOIs to all uploads. We created a “collection” for this conference\footnote{\url{https://zenodo.org/communities/coolstars20half}} and assigned an editor to approve uploads to the collection. Such a curated collection can be ingested into the ADS after the conference is over. That way, participants need to upload their poster only once, there is no second upload for proceedings later, although we leave time for the registrants who prefer to submit a paper style pdf (like this one) as opposed sharing a poster on ADS.  Zenodo provides more than enough space (50 GB per upload) and we do not have to worry about setting up or paying for hosting and databases. In general, that worked well. About 5\% of the participants contacted us with upload problems, usually because they forgot to press the “save” button, did not select the conference collection, or did not know that their posters would only appear in the collection or on our website after manual approval (i.e.\ with a time delay of a few hours even for the most diligent editor).  In general curation was straight-forward. During the final week, we spent a few minutes about 4 times a day checking the incoming posters.  Warning, if the editor does not keep up they will be faced with 300 posters to curate all at once.

Instructions on uploading cannot be too detailed. We provided over a dozen screenshots and several pages of text, and stuff still was missed.
We then used the Zenodo API to download a list of contributions in the collection, and display posters on our website (using a preview jpg/png if uploaded by the author to Zenodo or making a thumbnail from the poster pdf with a script), thus offering several different ways to see all posters:
\begin{itemize}
  \item On Zenodo (shows list with title, author, abstract)
  \item Our conference website (title, author, preview image, sorted by first keyword)
  \item The conference GatherTown (see section~\ref{sec:gathertown}; title, author, preview image, sorted randomly)
  \item In a movie (merged from all preview images)
\end{itemize}
A “view random poster” button on the website received especially good feedback, as it simulates walking through a poster hall until something catches your eye.

\subsection{Communication channels: Slack}
Slack worked great. We had one channel per plenary session (6 in total), one for posters, one for general announcements, and one per Topical Interest Room (TIR).
We also had one channel called \#hello-my-name-is. This channel was meant to cut down on “Hi”- chatter noise in the general channel and was not used as intended in the beginning. We should have explained the use of each channel when we initially  distributed the slack invite link. Beyond that participants helped each other out (e.g.\ with zoom problems). We also used a slack channel to collect questions in the zoom session, with the idea that discussion could continue there after a given talk ends, or even after the zoom closes. The latter did not happen much, so a single channel “\#plenary-questions” would have been sufficient.
The free version of slack keeps 10k messages. We ended with only 3.3 k used, so we conclude that the free version of slack is sufficient (slack can be quite pricey on a per participant basis, unless the hosting institution has a paid plan already).

Lessons:
Establish some slack-etiquette (greetings channel, reply in threads).  We also used slack to remind participants of the meeting code of conduct daily.  It mad a good forum for these reminders without implying any particular driver. 

\subsection{GatherTown}\label{sec:gathertown}
\begin{figure*}
	\centering
	\includegraphics[width=0.85\linewidth]{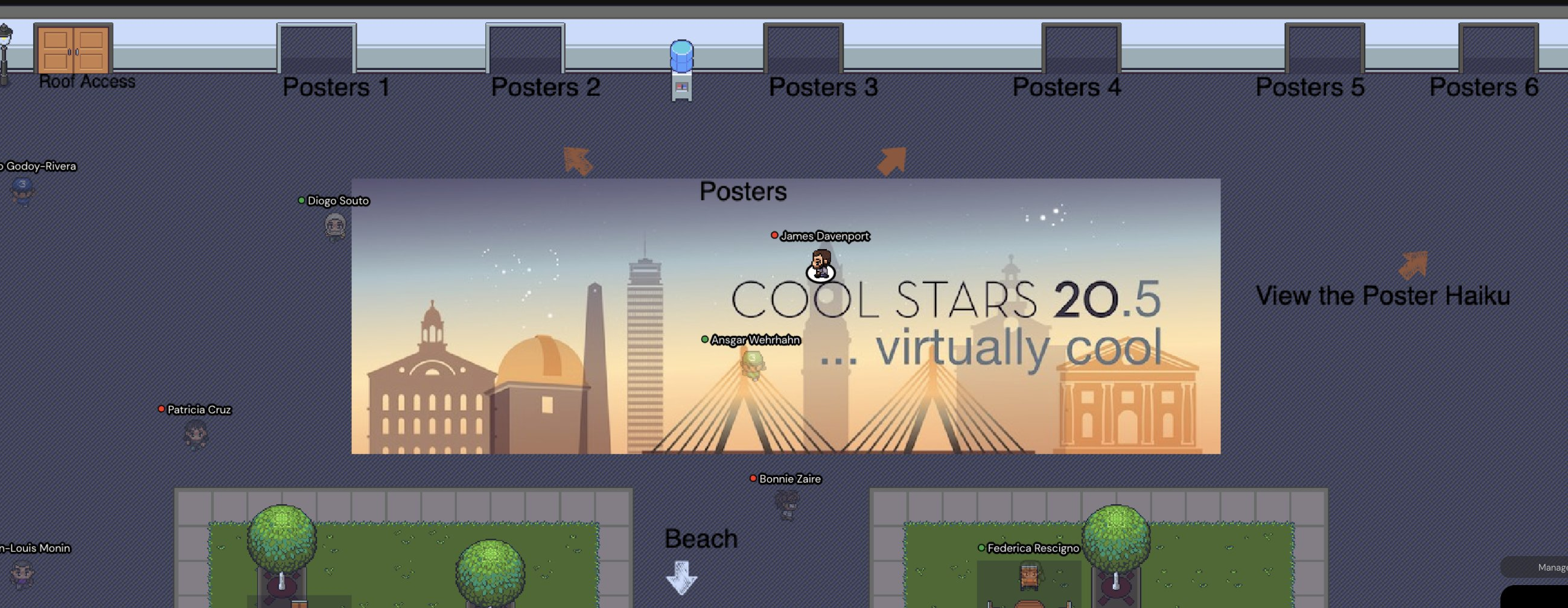}
	\caption{Screenshot of GatherTown. Avatars of different conference attendees are visible; when avatars get close to each other, a video chat automatically opens so you can ``hear'' people close to you. Cool stars 20.5 banner on the floor for decoration.}
	\label{fig:gathetown}
\end{figure*}
\url{https://gather.town/  }\ is a meeting platform with a retro video game feel, see Figure~\ref{fig:gathetown} for a screenshot. Participants connect via their web browser and walk around in virtual rooms. When two or more avatars are close to each other,  video/audio chat automatically connects. Similarly, poster thumbnails pop up when you approach them, allowing users to easily wander and browse the halls. Thus, compared with using Slack and Zoom alone, GatherTown gets closer to the feel of walking around real conference or poster halls and bumping into new people randomly or searching out friends and collaborators. GatherTown was an overwhelming success and received only positive feedback from those who used it. We recommend it as a social component for any online conference.
GatherTown is a commercial provider with pricing depending on the max number of people logged in at the same time, but offered us a deep academic discount (we paid \$600 for 24/7 open rooms with up to 300 participants using the highest server tier called ``metropolis'', which enabled unlimited posters).

Our GatherTown had six poster halls with 150 posters voluntarily contributed by attendees, and social spaces (a beach, a roof-top deck, private meeting rooms, games). Posters were submitted as jpeg or png images via email, and limited to file sizes of 3MB or less. We also made lower resolution ``thumbnail'' versions of each poster to make loading the proximity-based previews easier. Each poster was uploaded manually by the GatherTown admin.

GatherTown worked very well for those who tried it, with very few technical hiccups. Users of many ages/career stages enjoyed it.
We never had above 150 people in there, significantly less than the zoom plenary (300-500), perhaps becasuse it was too ``gimmicky'' for many, or they didn't wish to join an interactive session.
Poster rooms quickly filled up, but not everyone who submitted a poster to GatherTown attended. This is probably because we did not have a dedicated ``poster hour'', but ran GatherTown in parallel with our topical interest rooms and advertised it during ``coffee'' breaks.

We reached a ``critical mass'' of posters, such that people began assuming all posters were automatically in the session. This was particularly notable because we added directions for GatherTown posters only a couple days before the conference.
Many users wandered all the posters, viewing a very sizeable number (like at a big meeting). Though we only reached 150 people at once, there was a sustained group of at least 60-80 people present in GatherTown during the meeting hours.

The default poster room size is 24 posters, which we recommend using. Room designs are completely modifiable in GatherTown. We experimented making some larger poster rooms to accommodate  overflow, but this did not improve social engagement and added a bit of lag for users. 
We did not have exhibitor booths, but we note they work similar to posters and we would recommend them for larger virtual conferences.

We added some logos/art on the floor for decoration. This was well received.
The Haiku were available in a TV theater, but this was under used because it was not very obvious.
We had a number of private conference rooms or games rooms. Those were underused, though many users on Twitter enjoyed playing the games occasionally. The whiteboards were largely unused, since we did not hold splinter sessions within GatherTown.

People tweeted pictures of themselves on the ``beach'', which added to a real meeting feel. A roof deck-themed social room was the most popular hang-out space.  At one point a group formed a conga line and ``danced" to the beach.

Lessons:
\begin{itemize}
\item Keep the open area smaller to increase random interactions. Seeing people talk and interact close is a better motivation than empty hallways.
\item Private meeting rooms were not used widely, but having a few available is helpful (e.g.\ we had a Jobs room briefly).
\item People WILL use couches/chairs/benches. We anecdotally saw equal use of furniture for having discussions and for sitting idle (i.e.\ sitting on a bench, keeping Gather open in the background while in Zoom sessions).
\item Get clear instructions for GatherTown submission well ahead of time. This was not an option for us since we had issues with figuring out how to fund it.
\item Consider including ALL posters, and make GatherTown the default to view the posters.
\item Establish semi-strict resolution/file size requirements for the posters. 3 MB  for full posters, with a low-res thumbnail version of just the top quarter for the pop-up. This greatly helps latency during the poster session.
\item Have clear hours set aside for the GatherTown posters, which lasts at least one hour. Many users stuck around for 2 hours. One possibility in a world-wide meeting is to schedule social time an hour before the start of the plenary \emph{and} and hour after the last session of the day to accommodate participants in different time zones. We also used GatherTown for the reception, the night before the start of the meeting.
\item Make a map of the posters so people can look up posters, with links to the Zenodo version, too, if possible for future reference.
\item Keep the default 24 posters for rooms, remove unused posters. Utilize many rooms to keep per-room latency low.
\item Lock the Town with a password, and require users to use their full names. An Avatar marked ``anonymous'' or ``Rocket Man'' does not get interaction, since people are looking to interact with other people.
\item Have more than 1 admin who can load/fix posters. Only 1 or 2 poster upload errors were reported ($<1$\%).
\item Include a few games as ``easter eggs'', but don’t bother with a sophisticated game room.
\item Splinter sessions might work well in Gather Town. Plenary talks are probably best in Zoom, since the video compression is better and more stable, and Zoom allows recording. However, it is possible to run them both simultaneously if people have powerful enough computers.
\item Several attendees reported using their iPad for Zoom and a  laptop for GatherTown (or similar setups).
\item Use the room templates from GatherTown as much as possible. Much customization is possible, but the benefits likely do not outweigh the cost of time.
\end{itemize}

\subsection{Video-conferencing: Zoom}
Zoom worked well in general, but any hiccup in the live session quickly becomes a major problem, while a similar hiccup in anything that happens offline,  such as building a website, can be corrected without anyone noticing. A few specific lessons are listed below, intentionally kept short because features and software versions change so quickly:
\begin{itemize}
\item We used a session, not a webinar so in principle anyone can unmute, unless the host deactivates that feature.
\item Host should make certain everyone muted and camera off by default. In particular, people joining late listen to the talks and don’t check if they are auto-muted.
\item We made all speakers co-hosts, so that they could unmute and share their screen.
\item All speakers were requested to attended a practice/setup session an hour before the plenary started, most complied. This helped address issues like bad microphones and presenter display errors.
\item Sessions end when the host leaves. So, the host has to stay on and can’t quickly hang up to check another session. So you need a dedicated host for each session. 
\item We planned break-out sessions as individual meetings, but they were set up such that the same host had to be present in all of them simultaneously - which does not work. We then switched to using break-out rooms, but attendees with software version 5.3 or below (about 10\% of the participants) cannot enter break-out rooms on their own, they have to be assigned manually by the host (which requires some back-and-forth on chat and made every break-out start 10-15 min late).
\item A zoom session was set up for the wrong date and could not be joined, we had to slack/mass-email a new link, leading to some confusion.
\item Breakout rooms can exist without a host, but that means that nobody has the power to record the session or mute attendees with background-noise who forget to mute themselves.
\item You cannot leave a session (e.g.\ to go to a break-out room) if you are co-host.
\item Our biggest error was the expectation the primary host could record all sessions, including the simultaneous splinters.  This was an unjustified expectation.
\end{itemize}

\section{Lessons for organizing online plenaries}
In this section, we summarize the advice to running plenary sessions remotely (e.g.\ zoom). This is the most visible part, and, as we said above, planning is critical because there is little time to fix a mistake, we feel that this list deserves its own section.
\begin{itemize}
\item Practice, practice, practice. No matter how experienced your team, get together a group of at least ten people beforehand and test the exact same features that your meeting will have (parallel sessions, self-assigned break-out rooms, changing hosts, recording).
\item Share details beforehand in the organizing team, such that anyone in the team could start the session and claim host privileges, e.g. if power goes out somewhere (our SOC chair lost power at home during the meeting), a host has to be at multiple places at the same time, etc.
\item Establish non-zoom contacts (e.g.\ phone) between organizers.
\item Have several devices on hand, so that one host can log into several sessions at once if needed.
\end{itemize}

\subsection{Command center on the day of the conference}
Due to COVID, all organizers worked from home. In particular when things went wrong (e.g.\ a zoom link did not work), that means that communication was delayed because we had to try to fix it, but also email / slack with each other at the same time. Having the zoom guru, the SOC chair, and email/slack/website handler (in our case the LOC chair) in the same room would have helped and will be possible when virtual meetings are by choice and not due to a global pandemic.

\subsection{Other minor points}
\begin{itemize}
\item One of our speakers did not turn up. That happens in in-person conferences as well (e.g. missed flights), but as lesson for next time, we suggest to be extra-vigilant, looking out for the following warning signs: A speaker did not register, did not submit an abstract, did not update a title, did not communicate with the SOC in the four weeks before the conference.
\item Even in an online conference with no fee, some attendees still require a certificate of attendance (in our case 12/600). Not a big deal, but it helps to have a pdf with letterhead ready.
\end{itemize}

\subsection{Social media}
As with an in person conference, many participants tweeted thoughts and comments, this gave the meeting advertising ahead of time and a sense of persistence during. 

\begin{itemize}
\item We had a devoted account with multiple tweeters, we both created original tweets and responded to and retweeted others.
\item We had an advertised hashtag \texttt{\#cs20half}.
\end{itemize}
In general, this worked identically to an in person meeting.

Facebook was also used but was a less interactive format.

\bibliographystyle{cs20proc}
\bibliography{bib.bib}

\begin{thebibliography}{1}
\providecommand{\natexlab}[1]{#1}

\bibitem[\protect\astroncite{{Burtscher}
  \emph{et~al.}}{2020}]{2020NatAs...4..823B}
{Burtscher}, L., {Barret}, D., {Borkar}, A.~P., {Grinberg}, V., {Jahnke}, K.,
  \emph{et~al.} 2020, Nature Astronomy, 4, 823.

\end{thebibliography}

\end{document}